\documentclass[sigplan,nonacm]{acmart}

\usepackage{algorithm}
\usepackage{algpseudocode} 
\usepackage{multirow} 
\usepackage{array} 
\usepackage{subcaption}
\usepackage[dvipsnames]{xcolor}
\usepackage{bm}
\usepackage{enumitem} 
\usepackage{dblfloatfix}
\usepackage{hyperref}

\begin{document}
\title{Shift Parallelism: Low-Latency, High-Throughput LLM Inference for Dynamic Workloads}

\author{Mert Hidayetoglu, Aurick Qiao, Michael Wyatt, Jeff Rasley, Yuxiong He, and Samyam Rajbhandari   \\ Snowflake AI Research}

\begin{abstract}
Efficient parallelism is necessary for achieving low-latency, high-throughput inference with large language models (LLMs). Tensor parallelism (TP) is the state-of-the-art method for reducing LLM response latency, however GPU communications reduces combined token throughput. On the other hand, data parallelism (DP) obtains a higher throughput yet is slow in response latency. Best of both worlds does not exist, and it is not possible to combine TP and DP because of the KV cache variance across the parallelisms.

We notice Sequence Parallelism (SP---Ulysses in training) has similar properties as DP but with KV cache invariance. We adapt SP to inference, and combine it with TP to get the best of both worlds. Our solution: Shift Parallelism.


Shift Parallelism dynamically switches across TP and SP, and minimizes latency in low traffic without losing throughput in high traffic. The efficient GPU communications of Shift Parallelism yields up to i) 1.51$\times$ faster response in interactive workloads and ii) 50\% higher throughput in batch workloads, compared to a TP-only solution.

We evaluate Shift Parallelism with real-world production traces with dynamic traffic patterns as well as synthetic benchmarking patterns across models, context sizes, and arrival rates. All results affirm the same: Shift Parallelism has a better latency vs. throughput tradeoff than TP or DP, and hence obtains low latency without degrading throughput in dynamic workloads.


\end{abstract}

\maketitle 


\section{Introduction}

LLM inference has become the dominant workload in AI as its applications span agentic systems, chatbot (interactive) applications, model post-training (e.g. reinforcement learning), and image/video generation. The efficiency of inference systems is critical to both the performance and cost of AI applications. As of today, GPU parallelization is the prominent way of enabling large-scale AI production, and advanced multi-GPU parallelization techniques make nontrivial tradeoffs across key performance metrics.

The parallelism techniques for inference are largely inherited from training, yet inference is different from training in terms of workload characteristics. Training workloads are typically homogeneous, stable, and do not care about latency but only care about throughput.
Inference, on the contrary, is bursty and dynamic, and often has unpredictable traffic patterns.  Furthermore, different workloads have different performance requirements. As a result, leveraging parallelism techniques designed for training in inference results in complex performance and cost trade-offs.

\subsection*{Inference Workload Characteristics}

When people talk about inference systems, they often refer to interactive workloads, or batch workloads.

\textbf{\textit{Interactive workloads}} process requests with low concurrency to minimize the completion time of each request. The completion time latency is important when there is a chain of interactions between the user and the LLM, such as in REST applications. The completion latency depends on the time to first token (TTFT)---and time per output token (TPOT), which are both critical in real-time applications.
            
\textbf{\textit{Batch workloads}} involve a number of requests to be processed concurrently, and the latency of an individual request is not critical. For example, workloads such as batched summarization or translation of hundreds or thousands of documents can cause high-traffic bursts that require high combined throughput of input and output tokens to minimize the cost per token.

In enterprise systems, the request traffic pattern is often mixed and dynamically changes over time in unpredictable ways. In such dynamic settings, it is a challenge to optimize for different traffic patterns simultaneously, since existing parallelization techniques impose significant trade-offs.


\begin{figure}
    \centering
    \includegraphics[width=1\linewidth]{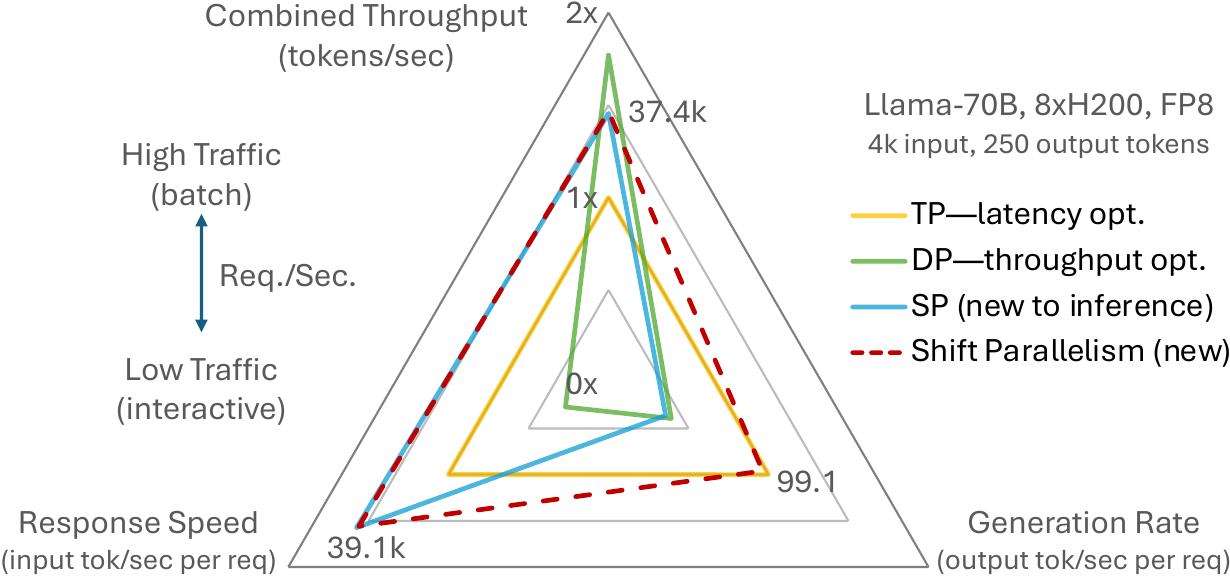}
    \caption{Comparison of response speed (\#input tok./TTFT) and generation rate (1/TPOT), and throughput (tokens/sec). Shift Parallelism obtains a higher throughput than TP in high traffic, and lower latency than TP and DP in low traffic.}
    \label{fig:trifecta}
\end{figure}

\subsection*{Latency vs. Throughput vs. Cost Tradeoff}
    
Existing parallelisms exhibit prohibitive latency vs. throughput (cost) trade-offs, as explained below.
    
\textbf{\textit{Tensor parallelism (TP)}} partitions the model weights and computation in each layer. It has to synchronize the embeddings across layers with costly all-reduce communications. By splitting model weights and computation across GPUs, it optimizes for latency (i.e., TTFT and TPOT), yet the communication overhead increases the cost (i.e., reduces throughput).

\textbf{\textit{Data parallelism (DP)}} parallelizes across request boundaries in embarrassingly parallel, providing high throughput. Yet, DP cannot speed up work within a single request, and therefore unsuitable for highly interactive workloads.

The first two rows of Table~\ref{tab:performance_tradeoffs} show the performance tradeoffs of TP and DP. We do have the choice to deploy TP and DP in separate nodes and route latency- and throughput-oriented requests, respectively. However, duplicating the node count (one for TP and one for DP) doubles the deployment cost and adds complexity.

\paragraph{Why can't we combine both?}
A performant and low-cost inference system should be able to switch between latency-oriented and throughput-oriented parallelisms swiftly in a single deployment based on traffic demands. But this is not viable with TP and DP because they have different attention layouts. Specifically, their KV cache memory layouts are incompatible, and switching requires complex and costly data movement.

However, in this work, we notice that  \textbf{\textit{Ulysses Sequence Parallelism (SP)}}~\cite{jacobs2023deepspeedulyssesoptimizationsenabling}---another form of parallelism developed and used in training---can offer a potential solution to resolving the challenge. SP splits the input sequence across GPUs to parallelize work within a single request to reduce TTFT. Unlike TP, it avoids costly all-reduce communication, while still achieving high GPU utilization. And while SP cannot parallelize decoding steps, resulting in the worst TPOT compared to TP and DP (Table~\ref{tab:performance_tradeoffs}), it has the same KV cache layout as TP, allowing for dynamic switching to TP when TPOT is critical. We call this dynamic approach  Shift Parallelism, which is the focus of this work.

\begin{table}[t]
    \centering
    \caption{Performance tradeoffs of inference parallelisms.}
    \begin{tabular}{c}
        \hspace{-1mm}\includegraphics[width=\linewidth]{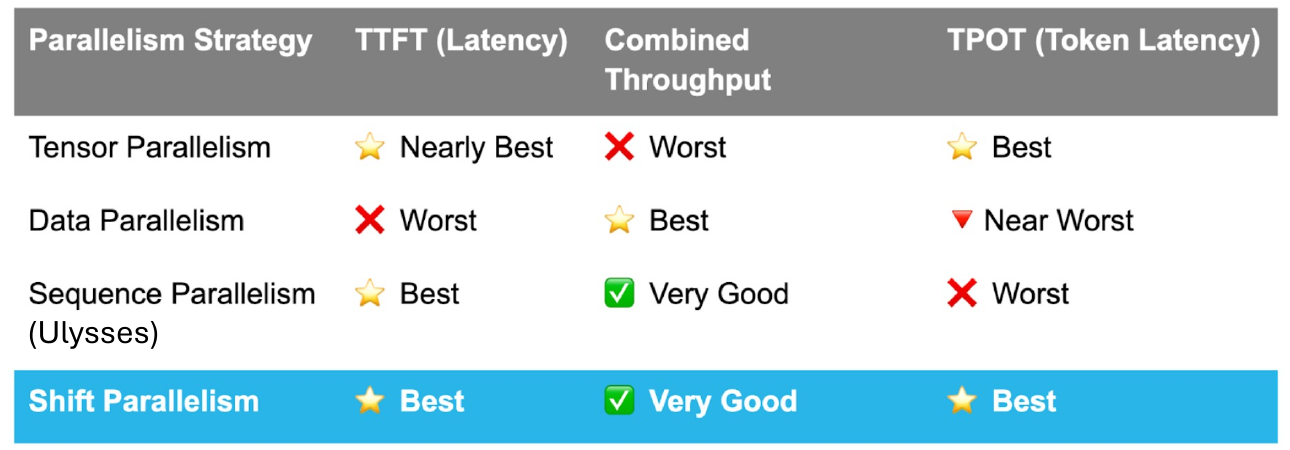}
    \end{tabular}
    \label{tab:performance_tradeoffs}
\end{table}

\textbf{\textit{Shift Parallelism}} dynamically chooses the parallelization strategy between SP and TP based on the real-world traffic pattern, identified by the number of batched tokens in each iteration. By a given threshold, Shift Parallelism uses:
\begin{itemize}
    \item TP for small batches---minimizing TPOT.
    \item SP for large batches---minimizing TTFT and achieving near-optimal throughput.
\end{itemize}
This is possible because the KV cache memory layout remains invariant between TP and SP, allowing Shift Parallelism to switch modes seamlessly, based on batch size and traffic patterns. More specifically, the KV cache layout does not change when switching across SP and TP. 

Figure~\ref{fig:trifecta} benchmarks the latency and throughput tradeoffs of related parallelisms. Shift parallelism provides 1.5$\times$ higher throughput than TP in high traffic and 1.5$\times$ faster response in low traffic, 2$\times$ faster generation than DP in low traffic while losing only 17\% throughput in high traffic.

In this paper, we
\begin{enumerate}
    \item We characterize inference workloads and identify latency vs. throughput tradeoffs with existing inference parallelization (TP, DP). 
    \item Adapt SP from training to inference and generalize it for a diverse set of inference models supporting GQA~\cite{ainslie2023gqa},  load-balancing at small batch sizes,  combination of TP, and KV cache replication when parallelism degree is higher than number of KV heads.
    \item Propose Shift Parallelism for dynamically switching across SP and TP for mitigating latency vs throughput tradeoff for dynamic workloads.
    \item Test Shift Parallelism with real-world production workloads and evaluate the performance characteristics via extensive benchmarking demonstrating up to 1.5$\times$ faster response time with 50\% throughput savings.
    \item Open source our implementation along with other SoTA techniques that are used in practice. 
    
\end{enumerate}

The rest of the paper is organized as follows. Section~\ref{sec:background} provides more details about production traffic patterns high-performance LLM inference. Section~\ref{sec:design} presents SP and Shift Parallelism. Section~\ref{sec:evaluation} evaluates the performance of proposed techniques on real-life patterns. Section~\ref{sec:related} about related work, and Section~\ref{sec:conclusion} concludes the paper.

\section{Background}\label{sec:background}

\begin{figure}[b]
    \centering
    \includegraphics[width=\linewidth]{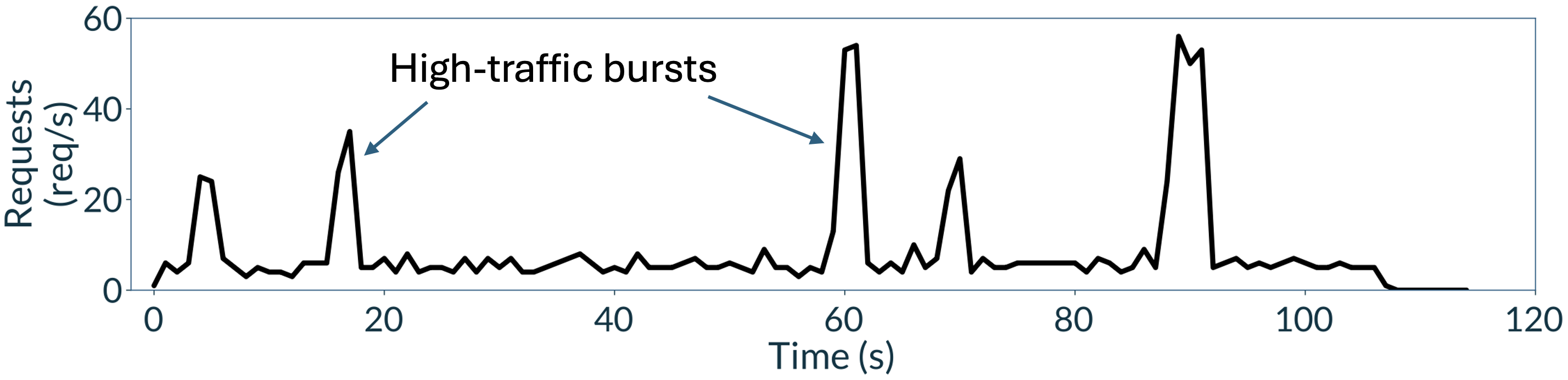}
    \caption{Bursty workload.}
    \label{fig:bursty_worload}
\end{figure}

\begin{figure*}[b]
  \centering
  \begin{subfigure}[b]{0.51\textwidth}
    \centering
    \includegraphics[width=\textwidth]{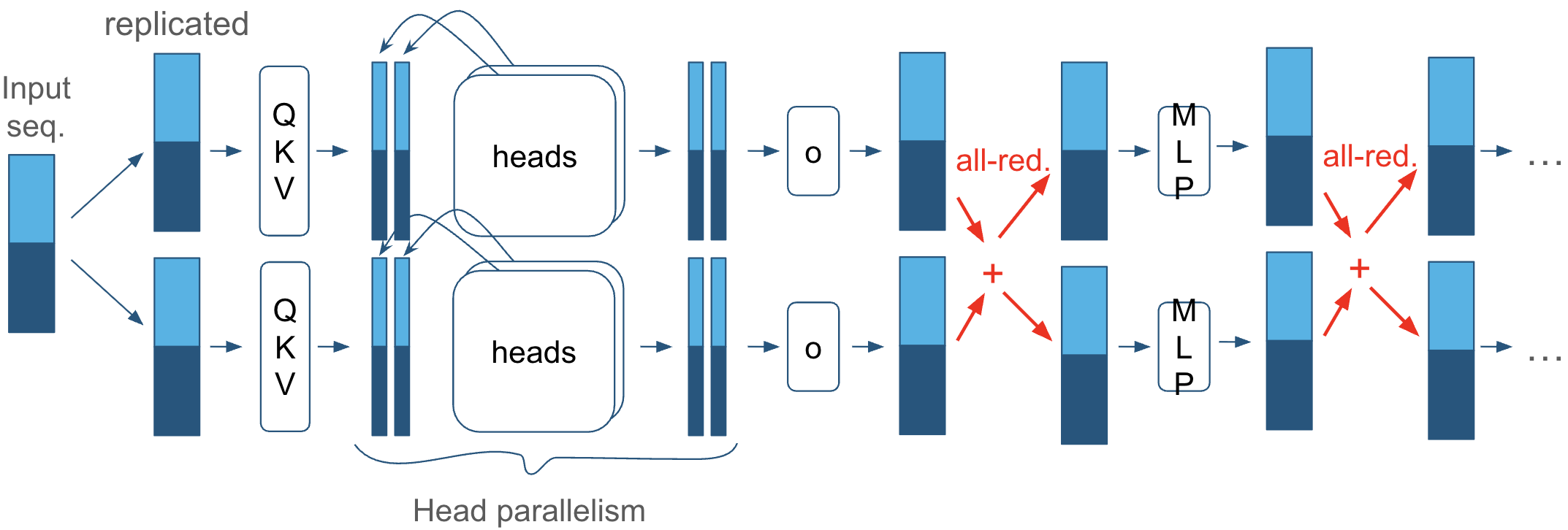}
    \caption{Tensor Parallelism (TP=2)}
    \label{fig:tensor_parallelism}
  \end{subfigure}
  \hfill
  \begin{subfigure}[b]{0.48\textwidth}
    \centering
    \includegraphics[width=\textwidth]{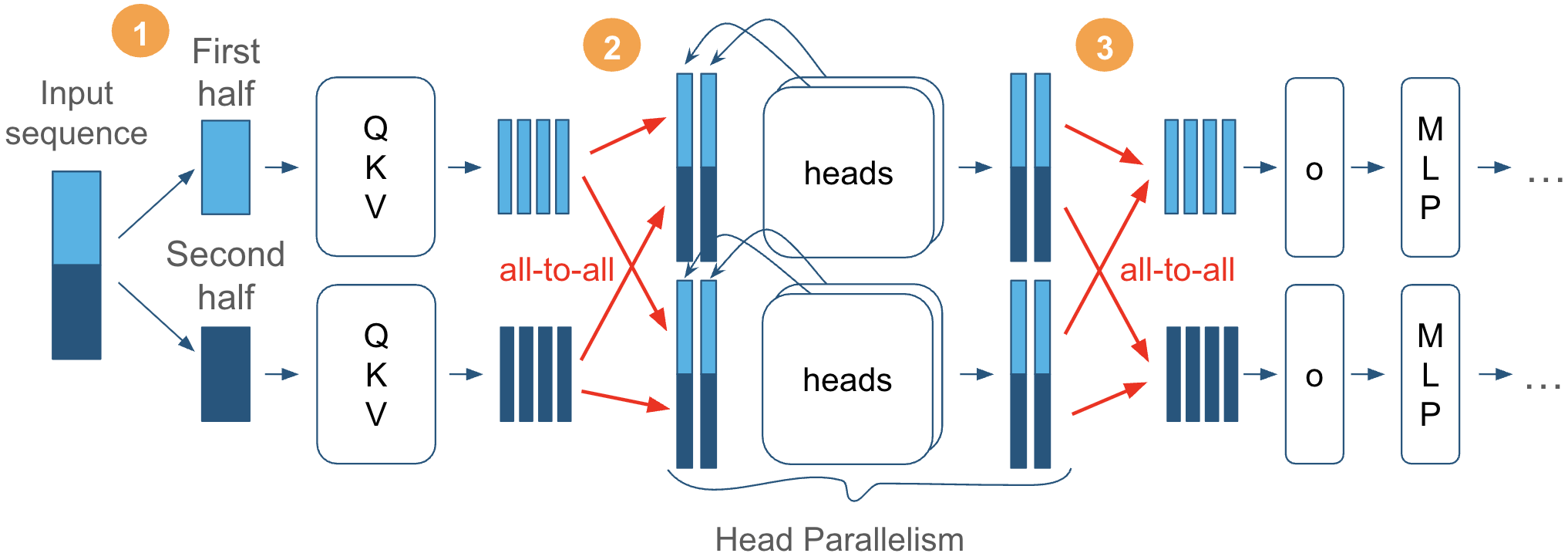}
    \caption{Ulysses Sequence Parallelism (SP=2)}
    \label{fig:sequcence_parallelism}
  \end{subfigure}
  \caption{Parallelization of the vanilla transformer on two GPUs with TP and SP. The attention has four heads which are parallelized across heads. In (b), SP (1) partitions the input sequence, (2) switches to head parallelism using an all-to-all communication, applies head parallelization to attention, and (3) returns back to SP.}
  \label{fig:combined}
\end{figure*}

\subsection{Production Traffic Patterns}

Today, LLMs are a foundational component of AI applications. A single deployment, such as Llama-3.3-70B, can serve diverse use cases including sentiment analysis~\cite{zhang2023sentimentanalysiseralarge,niimi2024dynamicsentimentanalysislocal}, retrieval-augmented generation (RAG)~\cite{lewis2021retrievalaugmentedgenerationknowledgeintensivenlp,gao2024retrievalaugmentedgenerationlargelanguage}, coding agents~\cite{zhang2024codeagentenhancingcodegeneration,jimenez2024swebenchlanguagemodelsresolve}, and more. These heterogeneous use cases produce dynamic traffic patterns that must be efficiently managed by the underlying infrastructure. For example, a coding agent typically issues a small number of repeated requests in a closed loop to iteratively refine its generated code, whereas a sentiment analysis workload may submit a large batch of requests in parallel to process text stored in a database.

In production, we typically observe two main \emph{classes} of requests. First, \emph{interactive} or \emph{latency-sensitive} requests (e.g., agentic or chatbot applications) generally arrive one or a few at a time, with response latencies—TTFT and TPOT (see Sec.~\ref{sec:performance-metrics})—directly shaping the user experience. Second, \emph{batch} or \emph{throughput-sensitive} requests usually arrive in large volumes (thousands to millions at once), where aggregate throughput (tokens/s) determines job completion time. When these two classes of workloads are mixed, the result is a highly \emph{bursty} traffic pattern, with different requests subject to different quality-of-service metrics (latency versus throughput). Fig.~\ref{fig:bursty_worload} illustrates an example traffic pattern that reflects our production environment.

\subsection{Performance Metrics}
\label{sec:performance-metrics}

Since LLM use cases are diverse, metrics that measure their inference performance are also multi-faceted. In our paper, we focus on three main metrics that cover the most important aspects of interactive and batch workloads:

\begin{itemize}
    \item \textbf{Time-to-first-token (TTFT, ms):} The time after a client submits a prompt until the first characters of response text (tokens) are received.
    \item \textbf{Time-per-output-token (TPOT, ms):} After the first response token is received, the time between each subsequent token until the response is completed.
    \item \textbf{Combined throughput (tokens/s):} The total number of tokens (both prompt and response) processed by the inference system per unit of time.
\end{itemize}

Typically, TTFT and TPOT shape the quality of service for interactive applications, while combined throughput shapes the quality of service for batch use cases and also impacts the cost of running the service for the model provider.

\subsection{Transformer Architecture}\label{sec:transformer}
\begin{figure}[t]
    \centering
    \includegraphics[width=\linewidth]{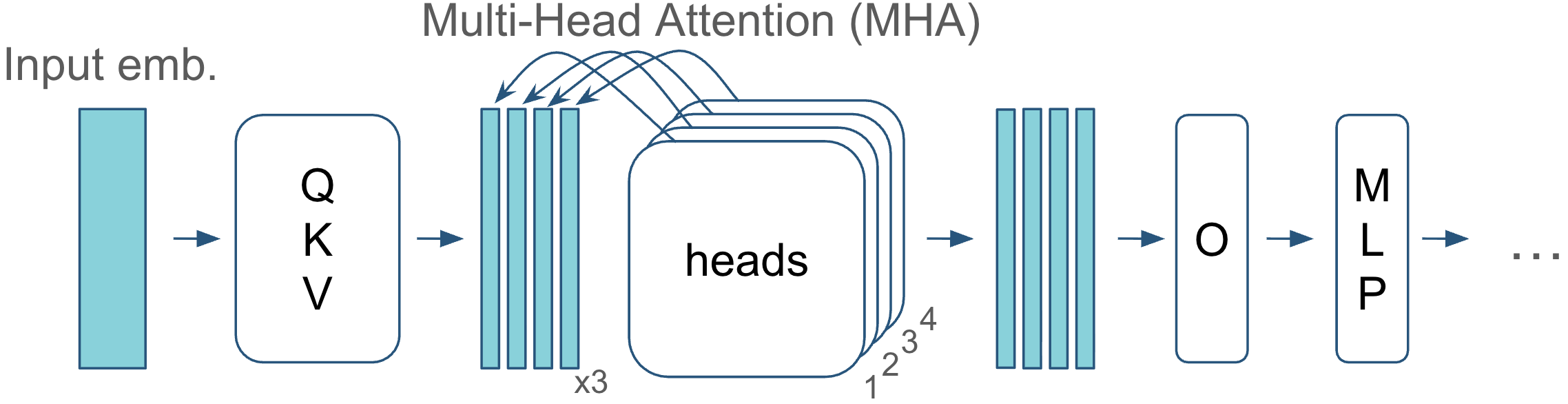}
    \caption{Vanilla transformer architecture and the attention mechanism.}
    \label{fig:transformer}
\end{figure}

A vanilla LLM involves a series of transformer layers, and each transformer layer consists of: i) an attention mechanism and a ii) multi-layer perceptron (MLP). The weights in the transformer layer correspond to the QKV (which is a concatenation of q---query, k---key, and v---value) and O matrices in the attention, as shown in Figure~\ref{fig:transformer}. 

First, the QKV matrix projects the input embeddings into the QKV space, where attention is applied. The Multi-Head Attention (MHA) consists multiple heads, each “attends” a different column of the input sequence. After the attention, the O matrix projects the attention output back to the embedding space to be further processed by the MLP layer.

Each LLM request involves a sequence of input and output tokens. In prefill, the input tokens are batched and propagated altogether over all of the transformer layers and initializes the KV cache for the attention layers. At the end of the prefill, the first output token is decoded according to the resulting probability distribution over all tokens in the vocabulary. For decoding the full output sequence, each new token is appended to the context sequence, and the attention patterns of subsequent context are reused from the KV cache.

\subsection{Existing Parallelism Approaches}\label{sec:existing_parallelism}



DP runs multiple replicas across requests and do not accelerate processing of a single request. For accelerating, TP partitions the weight matrices either row-wise or column-wise, as depicted in Figure~\ref{fig:tensor_parallelism}. Yet, row parallelization requires all-reduce with $O(n)$ communication cost, where $n$ is the sequence length. For a fixed sequence length, the communication-to-compute ratio increases with the TP degree as shown in the last column of Table~\ref{tab:complexity}.

\begin{table}[t]
    \centering
    \caption{Computational Complexity of TP and SP.}
    \begin{tabular}{c}
        \includegraphics[width=\linewidth]{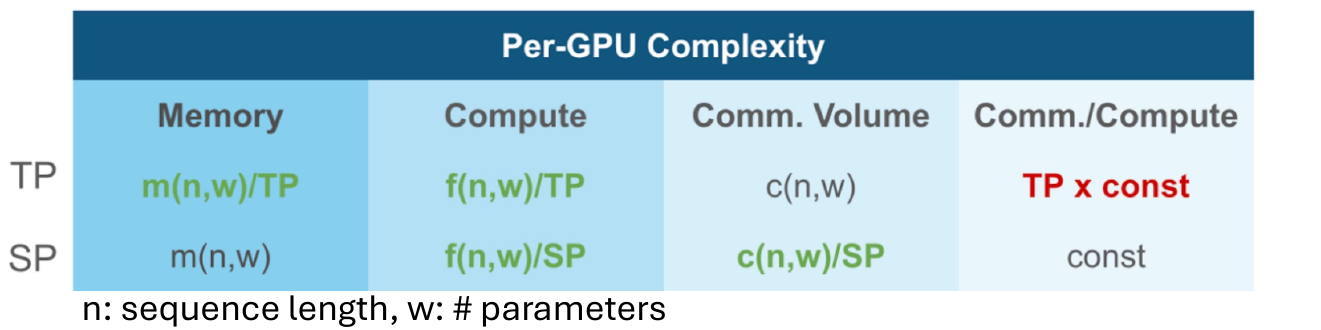}
    \end{tabular}
    \label{tab:complexity}
\end{table}

\emph{Head Parallelism} is commonly used with SP and TP, where the attention heads are distributed across the GPUs equally. This is done with no additional cost by column-wise partitioning of the QKV matrix, as Figure~\ref{fig:tensor_parallelism} shows. Head parallelism cannot be scaled beyond the number of heads.

\emph{Ulysses Sequence Parallelism} partitions the embedding sequence for parallelizing inference. Yet, each attention head requires the full sequence, resulting in all-to-all communications before and after the attention layer, as shown in Figure~\ref{fig:sequcence_parallelism}. Nevertheless, the communication cost does not increase with SP as shown in the last column of Table~\ref{tab:complexity}.

Ulysses has been applied to training, and in this work, we extend Ulysses for inference handling inference-specific nuances to allow for a generalized implementation \footnote{In the rest of the paper, we use Ulysses and SP interchangeably.}. 

\section{System Design and Implementation}\label{sec:design}
    
\subsection {Overview}





We design Shift Parallelism to enable switching between SP and TP, addressing the latency–throughput tradeoff in inference. The key insight is that both configurations must share the same KV cache layout—what we call KV cache invariance. This invariance allows us to switch seamlessly between SP and TP.

\begin{figure}[t]
    \centering
    \includegraphics[width=\linewidth]{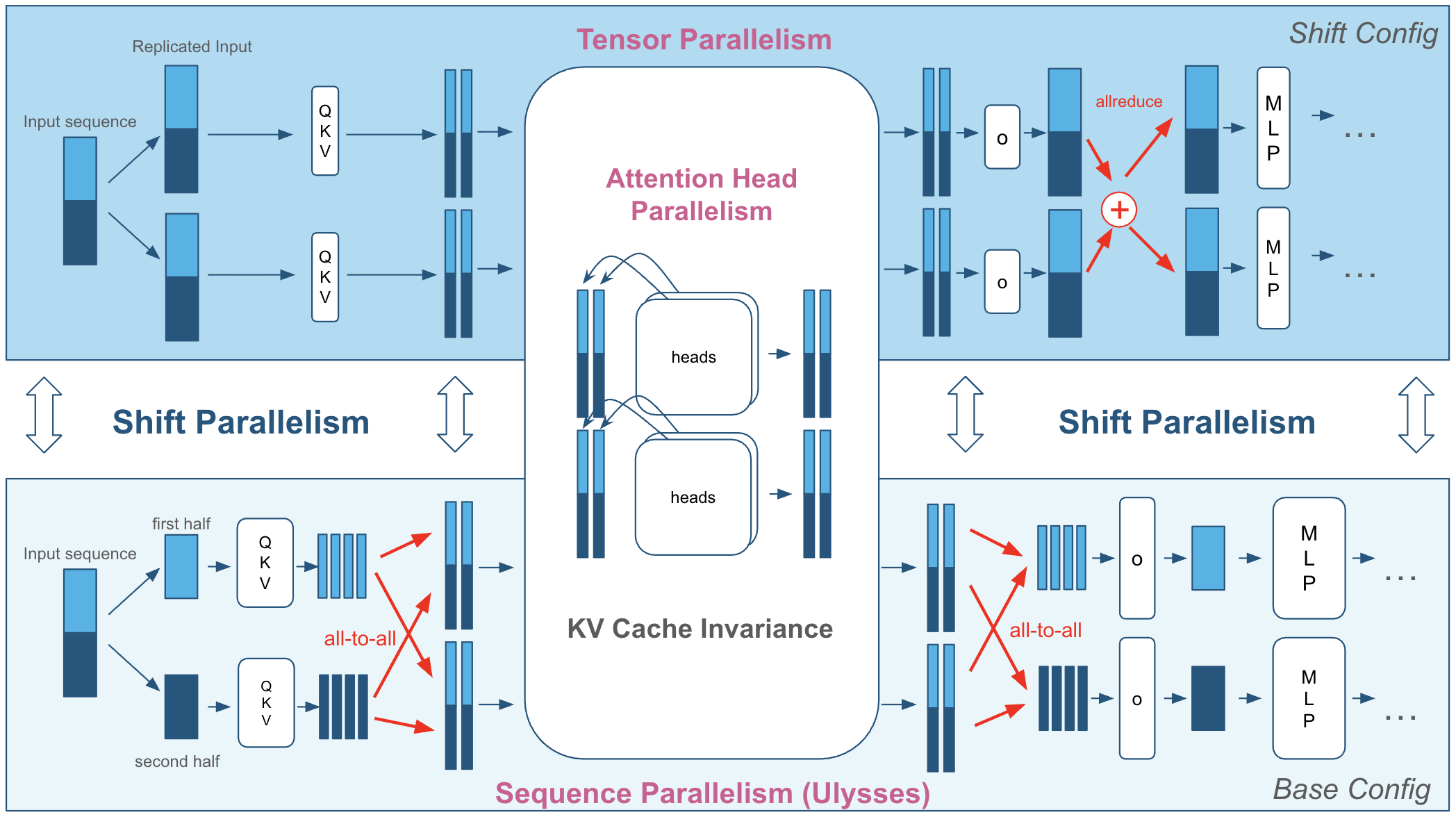}
    \caption{Although SP and TP are essentially different parallelisms, Shift Parallelism exploits the KV cache invariance between SP and TP for swiftly switching between them.}
    \label{fig:shift_parallelism}
\end{figure}

\textbf{Figure \ref{fig:shift_parallelism}} illustrates KV cache invariance between TP=2 and SP=2. In the center, four attention heads are evenly distributed across two GPUs (two heads per GPU). This distribution is identical under both $TP=2$ and $SP=2$, allowing the two configurations to share a single attention mechanism and KV cache.

\subsubsection{SP for Inference}

Applying SP to inference is more nuanced than in training because of variable traffic patterns (e.g., load imbalance) and the lack of Grouped Query Attention (GQA) support in earlier designs, and parallelism that can exceed the number of KV attention heads. To address this, we develop a fully generic SP for inference that: i) supports GQA, ii) replicates KV cache as needed, iii) handles load balancing under low-traffic scenarios.

Furthermore, real-world inference is not simply a choice between SP \textit{or} TP. For optimal performance, systems require arbitrary combinations of SP and TP. Our design supports this flexibility, enabling mixed $(SP, TP)$ configurations.


    


\subsubsection{Shift Parallelism}
Building on this flexible SP design, we implement Shift Parallelism using two configurations:

\begin{enumerate}
    \item \textit{Base configuration}: Uses either full SP or a mixed $(SP, TP)$ setup, as long as $SP \times TP = P$, where $P$ is the total number of GPUs in the node.
    \item \textit{Shift configuration}: Always $(SP=1, TP=P)$, spanning the full node.
\end{enumerate}

While conceptually simple, several technical challenges arise in enabling efficient transitions between these configurations:

\textbf{Cache invariance}\textbf{:} In general, the base and shift configurations are not automatically invariant. For arbitrary $(SP, TP)$ combinations, head ordering in head parallelism breaks the invariance. We resolve this by developing a general process-to-data mapping to ensure KV cache consistency in Section~\ref{sec:general_kv_cache}.

\textbf{Weight handling}\textbf{:}  Transitioning between configurations requires that weights be compatible across both base and shift modes. We consider two strategies: (i) on-the-fly slicing, and (ii) explicit weight replication. Based on memory cost analysis, we adopt the latter as the preferred approach discussed in Section~\ref{sec:implementation}.

We designed Shift Parallelism for ease of use and adoption. It is integrated into vLLM via a plug-in system (Section~\ref{sec:integration}) and is already deployed in production.


The rest of this section dives into the design details of SP and Shift Parallelism for inference.


\subsection{SP for Inference}\label{sec:sp_inference}

SP is essential for Shift Parallelism because it is the throughput-optimized counterpart of TP as they have the same KV cache layout, and therefore we can switch between them without changing the attention mechanism.

\subsubsection{Design for General Inference}


SP is originally implemented for training, and lacks important components, such as GQA mechanism~\cite{ainslie2023gqa}, that is commonly used in inference models. The original MHA mechanism is described in Section~\ref{sec:existing_parallelism}. 

\textbf{GQA Extension:} In this work, we extend SP for GQA mechanism for adapting SP to a diverse set of LLMs that are used in inference. GQA saves memory by sharing each KV head with multiple query heads. However, multi-GPU scaling of GQA is nontrivial with models involving small \#KV heads. For example, \texttt{Qwen-30B-A3B} attention has 4 KV heads, and it cannot be scaled to an 8$\times$GPU node since there are not enough KV heads to be distributed across more than 4 GPUs.

\textbf{KV Cache Replication:}\label{sec:kv_head_replication} TP solves the scaling problem by replicating the KV weights in the QKV projection (see Section~\ref{sec:transformer}), which recomputes the KV cache within GQA groups redundantly. This solution is not applicable to SP because each process owns only a slice of the input sequence and the missing slices cannot be replicated simply by recomputation.

For SP in inference, we implemented the KV cache replication using all-to-all communications. The KV heads are replicated within the send buffers of the collective call, resulting in replication in the receiving buffers across GPUs.


\textbf{Fusing Communications:} The QKV matrix fuses operations related to q, k, and v, bringing multiple communications down to a single matrix all-to-all communications together as represented in Algorithm~\ref{alg:combined_SP_TP} Line 4.

The GQA implementation replaces $3\times h$ with $h + 2\times h_{kv}$, where $h$ is the \# q heads and $h_{kv}$ is the \# k and v heads, and replicates the KV cache if necessary for handling any combination of $h$ and $h_{kv}$ with a single, e.g., fused, all-to-all communications.




\textbf{Small Batch Size and Load Imbalance:}\label{sec:load_imbalance} The main problem with SP is the load imbalance with small batch sizes, i.e., $n$ is small in Algorithm~\ref{alg:combined_SP_TP}. This problem does not exist in training because batches are large and static, whereas in inference, the batch size varies according to the traffic.

Specifically, decoding in low traffic yields small batch sizes because there are only a few tokens produced at a time. Small batch sizes comparable to the SP degree cannot be evenly partitioned across GPUs, causing serious load imbalance. For example, when the batch size is 9 and $SP=8$, all GPUs will process a single token except the one that processes two tokens, causing 50\% efficiency. SP even breaks down when $SP > \textrm{batch size}$, causing sparse communications.

To provide load balancing, we pad batches up to a multiple of SP degree so that we can evenly distribute them. Nevertheless, the padding results in redundant tokens, yielding a longer TPOT and hence longer request completion time compared to TP in low-traffic decoding.










\subsubsection{Combined $\boldsymbol{(SP, TP)}$ Algorithm}\label{sec:SP_TP_algorithm}

We need to combine SP with TP for handling large models that do not fit (or barely fits) in a single GPU. For a throughput-optimal config, we avoid partitioning the model with TP as much as each partition fits into GPU memory, and there is enough room for KV cache for providing concurrency and high throughput. Then the rest of the GPUs can be efficiently employed using SP, which enlarges KV cache. For example, our evaluation involves \texttt{Llama-17B-16E} (FP8) has 109 GB memory footprint, yet needs at least $TP=2$ for processing long contexts concurrently within 141 GB GPU memory, and therefore we need a combination of $(TP=2, SP=4)$ for an optimal deployment on a node with 8 GPUs.

\begin{algorithm}[t]
\footnotesize
\caption{Combined $(\textcolor{ForestGreen}{SP}, \textcolor{BrickRed}{TP})$ for the base config.}
\label{alg:combined_SP_TP}
\begin{algorithmic}[1]
\State $embed[n/\textcolor{ForestGreen}{SP}, d] \gets \textcolor{ForestGreen}{SP}.slice(input\_embeds[n,d])$
\For{$i = 1, \dots, L$}
    \State $qkv\_heads[n/\textcolor{ForestGreen}{SP}, 3 \times h / \textcolor{BrickRed}{TP}] \gets 
    embed * layer_i.qkv[d, 3 \times h / \textcolor{BrickRed}{TP}]$
    \State $qkv\_heads[n, 3\times h / (\textcolor{ForestGreen}{SP} \times \textcolor{BrickRed}{TP})] \gets 
    \bm{\textcolor{ForestGreen}{SP}.all\_to\_all}(qkv\_heads)$
    \State \fbox{$attn\_o[n, h/ (\textcolor{ForestGreen}{SP} \times \textcolor{BrickRed}{TP})] \gets layer_i.attn(qkv\_heads)$}
    \State $attn\_o[n/\textcolor{ForestGreen}{SP}, h/\textcolor{BrickRed}{TP}] \gets \bm{\textcolor{ForestGreen}{SP}.all\_to\_all}(attn\_o)$
    \State $embed[n/\textcolor{ForestGreen}{SP}, d] \gets attn\_o * layer_i.o[h/\textcolor{BrickRed}{TP}, d]$
    \State $\bm{\textcolor{BrickRed}{TP}.all\_reduce}(embed)$
    \State $act[n/\textcolor{ForestGreen}{SP}, d'/\textcolor{BrickRed}{TP}] \gets embed * layer_i.mlp\_up[d, d'/\textcolor{BrickRed}{TP}]$
    \State $embed[n/\textcolor{ForestGreen}{SP}, d] \gets act * layer_i.mlp\_down[d'/\textcolor{BrickRed}{TP}, d]$
    \State $\bm{\textcolor{BrickRed}{TP}.all\_reduce}(embed)$
\EndFor
\State  $output\_embeds[n, d] \gets \bm{\textcolor{ForestGreen}{SP}.all\_gather}(embed[n/\textcolor{ForestGreen}{SP},d])$
\State \Return $output\_embeds$
\end{algorithmic}
\end{algorithm}

Algorithm~\ref{alg:combined_SP_TP} shows the forward pass algorithm with an arbitrary $(SP, TP)$ configuration, where $n$ is the sequence length, $d$ is the hidden dimension, and $h$ is the number of heads. 

\subsection{Shift Parallelism}
Shift parallelism (the main contribution of this paper) is designed to obtain low latency (TTFT and TPOT) in low traffic, and high throughput in high traffic by switching across parallelisms.
The optimal parallelisms that we cover with Shift Parallelism are summarized in Table~\ref{tab:optimal_configurations}.

\begin{table}[b]
\small
\centering
\caption{Optimal Parallelisms Covered by Shift Parallelism.}
\label{tab:optimal_configurations}
\begin{tabular}{lcc}
& \textbf{Low Traffic}   & \textbf{High Traffic}        \\ \cline{2-3} 
\multicolumn{1}{l|}{\textbf{TTFT}} & SP                      & \multicolumn{1}{c|}{SP}     \\
\multicolumn{1}{l|}{\textbf{TPOT}} & TP                      & \multicolumn{1}{c|}{SP}     \\ \cline{3-3}
\textbf{Throughput}                & \multicolumn{1}{|c|}{SP* or TP} & \multicolumn{1}{c}{DP} \\ \cline{2-2} 
\end{tabular}\\ \vspace{1mm}
*SP for long input, TP for long output.
\end{table}


We can apply shift parallelism only across TP and SP because of their KV cache invariance property (Section~\ref{sec:general_kv_cache}), as a result, Shift Parallelism provides superior performance in the highlighted cases. The only case Shift Parallelism loses on DP (but wins on TP) is the throughput in high traffic, because parallel attention inevitably requires GPU communications.

In shift parallelism, we have two configurations; a) the base config that implements SP to optimize TTFT and throughput, and b) the shift configuration that implements full TP to optimize TPOT. The base configuration can optionally be a combination of TP and SP (Section~\ref{sec:SP_TP_algorithm}), if the model does not fit into a single GPU.

\paragraph{How do we shift?} The main criterion of switching between configurations is simple: We choose the base model for large batch size and the shift model for small batch size. Therefore, we decide on a shift parallelism threshold, if the batch size is larger than the threshold, we choose $(SP, TP)$ configuration, and choose the shift  configuration, i.e., full-TP on $(SP\times TP)$ group as Algorithm~\ref{alg:shift_SP_TP} describes.

\begin{algorithm}[h]
\footnotesize
\caption{Shift parallel $(\textcolor{ForestGreen}{SP} \times \textcolor{BrickRed}{TP})$ forward pass.}
\label{alg:shift_SP_TP}
\begin{algorithmic}[1]
\If{$n > threshold$}
    \State \Return {Algorithm \ref{alg:combined_SP_TP}[$\textcolor{ForestGreen}{SP}, \textcolor{BrickRed}{TP}$]($input\_embed[n,d]$)}
\Else
    \State \Return {Algorithm \ref{alg:combined_SP_TP}[$1, \textcolor{ForestGreen}{SP} \times \textcolor{BrickRed}{TP}$]($input\_embed[n,d]$)}
\EndIf
\end{algorithmic}
\end{algorithm}

\subsubsection{General KV Cache Invariance}\label{sec:general_kv_cache}

The KV cache invariance does not only require the same attention head layouts, but also the same ordering of the heads. Interestingly, the invariance across SP and TP breaks down when shifting across arbitrary $(SP, TP)$ and $(SP\times TP)$ configurations. When the base configuration involves a combination of SP and TP, e.g., $(SP=3, TP=2)$ the attention head order does not follow the same order as in $TP=6$, i.e., not $(0, 1, 2, 3, 4, 5)$, anymore but $(0, 2, 4, 1, 3, 5)$. The attention mechanism does not care about the order of the heads as long as we lock in to an order (i.e., base config's) and stick to it when switching to the shift config as depicted in Figure~\ref{fig:kv_invariance}. 

Figure~\ref{fig:kv_invariance} shows the distributed memory layout of Q projection with the (a) base config and (b) the shift config. The original config yields  TP groups $(0,1), (2, 3), (4, 5)$ and SP groups $(0, 2, 4), (1, 2, 5)$. TP groups partition the Q weights across heads (i.e. columns) and replicate the input embeddings. SP groups partition the embeddings across the sequence (i.e., rows), and replicate the Q weights. As a result of the 2D partitioning, the output of the linear layer ($q\_$) has a global layout as shown in the figure. As a result of the all-to-all communication within the SP groups, the head partitions ($q$) have interleaved head ordering $(0, 2, 4, 1, 3, 5)$. We need to adjust the attention head ordering of the shift config accordingly to provide KV cache consistency.

\subsubsection{Memory Management}\label{sec:implementation}

There are two ways of implementing Shift Parallelism with generalized KV cache invariance. 1) slicing the model weights on-the-fly and 2) loading separate models that share the attention mechanism (and the KV cache). We use 2) in our implementation.

\paragraph{On-the-fly slicing}

\begin{figure}[t]
    \centering    \includegraphics[width=0.92\linewidth]{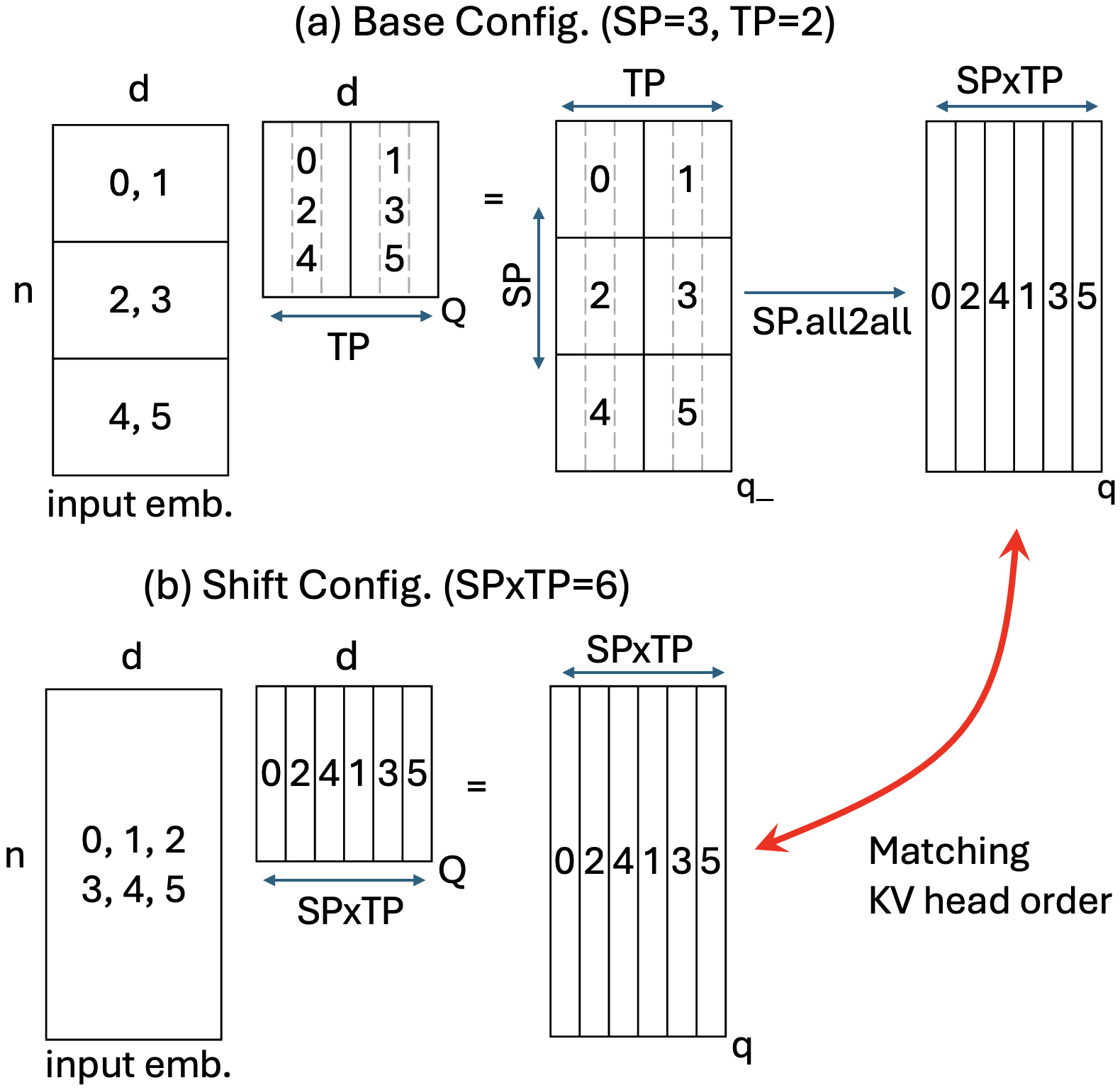}
    \caption{KV cache invariance of six heads across the base and shift configs. The shift config should shard the Q weights according to SP and TP degrees of the base config.}
    \label{fig:kv_invariance}
\end{figure}

This implementation modifies the linear layer implementation of the original code such that each GPU multiplies a slice of the base model's weight partition. To preserve KV cache invariance, each GPU must have the slice according to their SP ranks. For example, global ranks 2, and 3 in Figure~\ref{fig:kv_invariance} gets heads 1, and 4 which are already in the base model's respective weight partition.

Slicing provides the same effect with TP, and has no memory overhead since the running buffer can be reused for all layers. Nevertheless, it is not as performant as the next solution because each slicing requires matrix transposition due to an FP8 hardware limitation of Hopper tensor cores.

\paragraph{Separate Models}

The separate model solution does not share the weights across the base and shift configs, but replicates the weights. In this implementation, we load two separate models, one for the base configuration and one for the shift configuration, and these models share the same KV cache.

When loading the weights for the shift model, we use a separate group, \texttt{SP\_TP} that spans both \texttt{SP} and \texttt{TP} groups, but with the order of \texttt{SP} group to preserve KV cache coherency. As a result, the shift model will load the right weight shards as shown in Figure~\ref{fig:kv_invariance}.
\begin{itemize}
    \small
    \item \texttt{TP: [[0, 1], [2, 3], [4, 5]]}
    \item \texttt{SP: [[0, 2, 4], [1, 3, 5]]}
    \item \texttt{SP\_TP: [[0, 2, 4, 1, 3, 5]]}
\end{itemize}

With the separate model solution, we can write the total weight footprint as
\begin{equation}
    w_{total} = \frac{w_{base}}{TP} + \frac{w_{base}}{SP\times TP},
\end{equation}
where the first and second terms represent the base and shift models' weights, respectively. As a results, the memory overhead of the shift model is $1/SP$, i.e., a base model with more $SP$ and less $TP$ alleviates the memory overhead of the shift model. For example, when $SP=8$, the shift model's memory overhead is 12.5\%.




\subsection{Integration into vLLM}\label{sec:integration}

None of the existing inference frameworks implement SP (Ulysses), and also modifying existing enterprise frameworks (such as vLLM) is tedious. We overcame the problem by developing a plug-in system~\cite{arcticInferenceRepo}
for implementing the proposed techniques in this paper.

For achieving low latency in inference, it is crucial to enable compilation and CUDA graph capture mechanisms in vLLM. The plug-in system compiles and captures both base model and shift model separately. Capturing separate graphs for multiple shapes yields hundreds of graphs, which are registered during initialization and replayed accordingly at runtime. The additional graphs for the shift model do not increase the capturing time or memory significantly.

\section{Evaluation}\label{sec:evaluation}

In this section, we demonstrate that Shift-Parallelism can mitigate the latency vs throughput tradeoffs commonly seen in TP and DP. More specifically, we show  
\begin{enumerate}[left=0pt]
    \item  Shift Parallelism can adapt to bursty synthetic traffic pattern, achieving simultaneously lowest latency (up to 3.23$\times$ lower) and near optimal throughput compared to TP and DP.
    \item On open-source production traces, Shift Parallelism: 
    \begin{itemize}
        \item Consistently obtains the lowest TTFT and TPOT, and hence completion time statistics compared to TP and DP when running Azure LLM Code Trace ~\cite{patel2024splitwise}.
        \item Can keep up with the request traffic with no wait time whereas TP and DP cause growing wait times when running Mooncakce Conversation Trace~\cite{qin2025mooncake}.
    \end{itemize}
    \item Extensive evaluation of Shift Parallelism over a wide range of sequences and request traffic demonstrating consistently superior performance (1.67$\times$--6.97$\times$ faster response, 1$\times$--2.45$\times$ faster generation and 1.51$\times$ higher throughput) compared to TP and DP, guaranteeing lowest latency with low cost over the entire spectrum, even in high traffic.
    \item Shift Parallelism can accelerate real-world production deployment offering fastest open-source inference solution (3.4$\times$ lower completion time, and 1.06$\times$ higher throughput) by composing with SoTA inference technologies like SwiftKV and Speculative Decoding,
    \item The cost breakdown analysis of DP, TP, and Shift Parallelism and explore further tradeoffs across dense and sparse models.
\end{enumerate}

\subsection{Experimental Setup}

\subsubsection{Hardware}\label{sec:hardware} Unless specified, we use AWS instances with 8xH200 GPUs each, i.e., \texttt{p5en.48xlarge}. Each GPU has 141 GB memory with 4.8 TB/s bandwidth, and also provides a peak dense matrix multiplication of 1,979 FP8 TFLOPS with tensor cores. The GPUs are interconnected with an NVSwitch network with 900 GB/s rated bandwidth.

\subsubsection{Software}
Unless specified, we use our implementation (Sec.~\ref{sec:integration}) plugged into vLLM v0.9.2. For comparison, we use SGLang~\cite{zheng2024sglangefficientexecutionstructured} v0.4.6, TRT-LLM~\cite{nvidia_tensorrt_llm_2023} v0.18.2. 

\begin{table}[h]
\footnotesize
\centering
\caption{Models used in evaluation.}
\label{tab:models}
\begin{tabular}{r|ccccc}
 & \textbf{Num.} & \textbf{Num.} & \textbf{Hidden} & \multicolumn{2}{c}{\textbf{\# Heads}} \\
\textbf{Model Name} & \textbf{Params.} & \textbf{Lay.} & \textbf{Size} & \textbf{Q} & \textbf{KV} \\ \hline
\texttt{Llama-70B} & 70B & 80 & 8192 & 64 & 8 \\
\texttt{Qwen-32B} & 32B & 64 & 5120 & 64 & 8 \\
\texttt{Llama-17B-16E} & 109B/17B & 48 & 5120 & 40 & 8 \\
\texttt{Qwen-30B-A3B} & 30B/3B & 48 & 2048 & 32 & 4 \\
\end{tabular}
\end{table}

\subsubsection{Models}

We use the models listed in Table~\ref{tab:models}, all with FP8 quantization. Shift Parallelism is originally designed for dense models, therefore we first present the main evaluation for \texttt{L70B} and \texttt{Q32B}, and then we discuss the performance limitations with mixture of experts (MoE) models---\texttt{Q30B-A3B} and \texttt{L17B-16E}---their static and active number of parameters are shown separately in Table~\ref{tab:models}.

\begin{figure}[b]
    \centering
    \includegraphics[width=\linewidth]{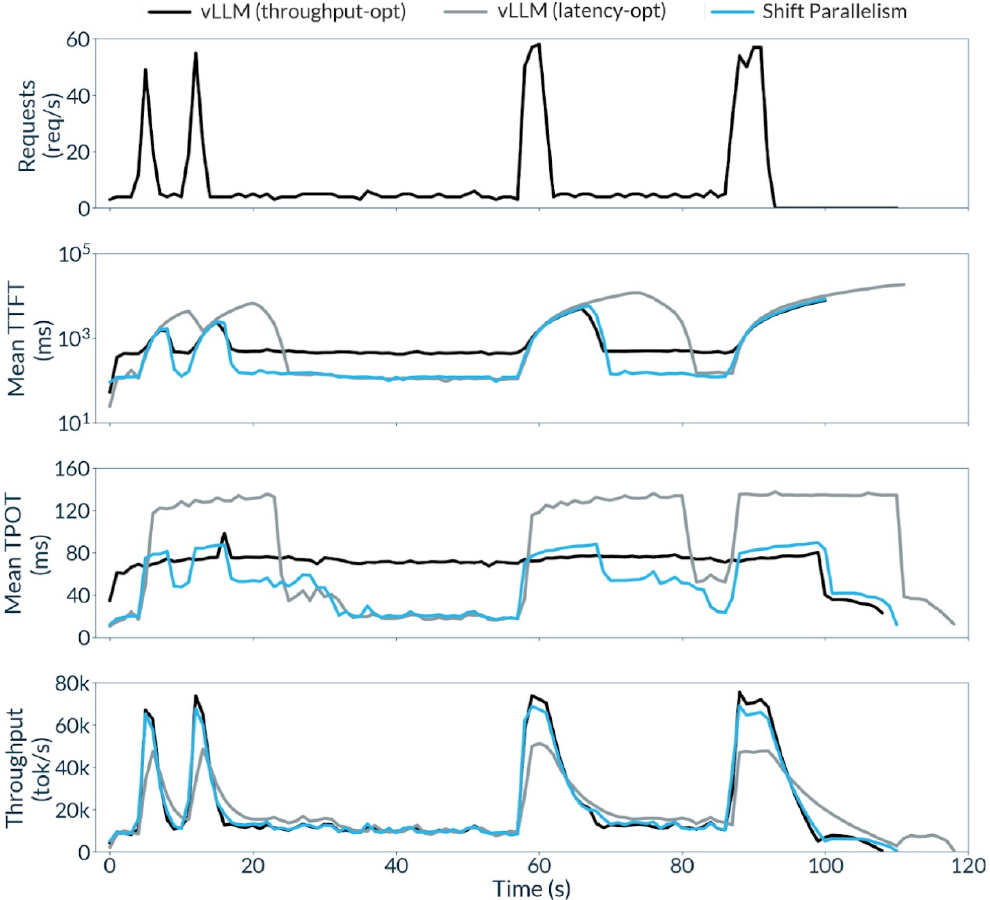}
    \caption{Shift Parallelism achieves the lowest response, fastest generation and near-optimal throughput under dynamic traffic. We used Llama-70B and a modified version vLLM's serving benchmark that makes requests a steady stream of request at low frequency with occasional bursts of high frequency requests.}
    \label{fig:bursty_traffic_result}
\end{figure}

\subsubsection{Datasets}\label{sec:datasets}
We use the following datasets in the order of presentation. i) A bursty synthetic pattern that resembles real-life production environment, ii) a couple of open-source production traces (details are given in Section~\ref{sec:real_world_datasets}), iii) synthetic requests with random data for parameterized benchmarking, and iv) a trace that is a mixture of requests from HumanEval~\cite{chen2021evaluatinglargelanguagemodels} and from a CodeAct agent~\cite{neubig2024-openhands-codeact-2.1:} running against SWEBench~\cite{jimenez2024swebenchlanguagemodelsresolve} (HumenEval is one-shot and SWEBench is agentic). We filter these requests to create a synthetic dataset requests for the sake of running speculative decoding in Section~\ref{sec:comparison}.

\subsection{Latency and Throughput in Real-World Traffic }\label{sec:real_life}

\subsubsection{Bursty Synthetic Workload}

For testing Shift Parallelism on real-life environment, we create a bursty dataset by changing the arrival using vLLM's burstiness benchmark.
Figure~\ref{fig:bursty_traffic_result} (top) shows resulting traffic pattern that has four high-traffic bursts. The rest of the results show the input latency (i.e., TTFT) and the output latency (i.e., TPOT) that is experienced by a request in milliseconds, and also the combined input/output token throughput of all requests in tokens per second.

To obtain Figure~\ref{fig:bursty_traffic_result}, we randomly mix two real-life datasets that are described in Section~\ref{sec:datasets}. The mix involves both latency- and throughput-critical requests with variable sizes.

Table~\ref{tab:bursty_traffic_result} summarizes the latency and throughput statistics that are collected from the trace of the bursty workload experiment (Figure~\ref{fig:bursty_traffic_result}). The experiment trace with vLLM's TP and DP, and also the proposed Shift Parallelism shows

\begin{itemize}[left=0pt]
    \item Shift Parallelism obtains the lowest latency across TP and DP with bursty (dynamic) traffic since it can sustain low latency with a higher traffic than the other two .
    \item Shift Parallelism obtains a higher peak throughput than TP, and therefore processes the batches in a shorter time. As a result, the wait time for latency-critical request is reduced significantly, i.e., TTFT does not explode with Shift Parallelism (148 ms vs. 3.9 sec.).
\end{itemize}

\begin{table}[h]
\footnotesize
\centering
\caption{Performance stats with the bursty workload.}
\label{tab:bursty_traffic_result}
\begin{tabular}{l|lll}
 & Median & Median & Peak \\
 & TTFT & TPOT & Throughput \\ \hline
vLLM (throughput opt.---DP) & 1,355 ms & 83 ms & 75,535 tok/s \\
vLLM (latency opt.---TP) & 3,930 ms & 85 ms & 51,162 tok/s \\
vLLM+Shift Parallelism & 148 ms & 51 ms & 69,147 tok/s
\end{tabular}
\end{table}

Overall, Shift Parallelism can handle the high-traffic bursts better than both TP ad DP, achieving up to 9.16$\times$ lower TTFT and 1.63$\times$ lower TPOT than both, while at the same time achieving nearly as good throughput as DP during  high-traffic periods, ultimately, improving the quality of service.

\begin{figure}[t]
    \centering
    \includegraphics[width=\linewidth]{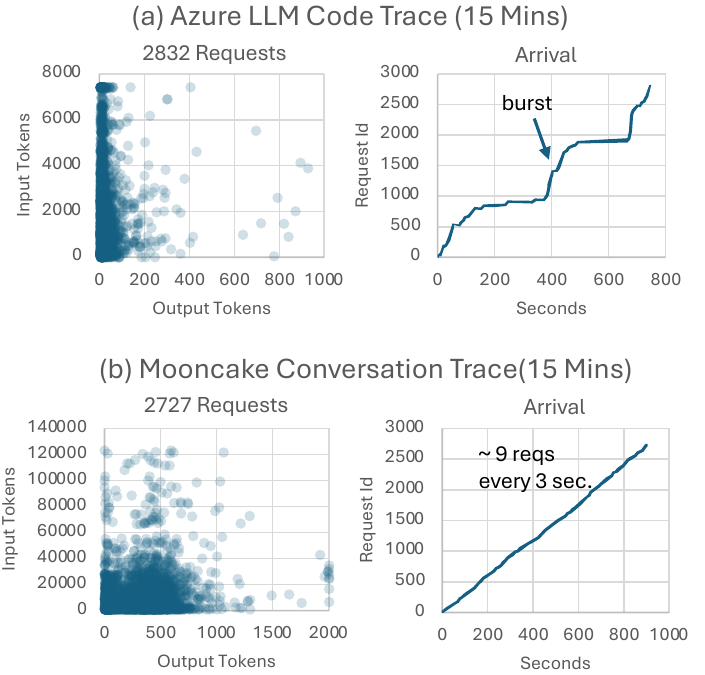}
    \caption{Input/output distribution and arrival times of real-world requests. (a) represents a bursty workload due to agentic code completion, causing low-traffic (silent) and high-traffic (burst) regions. (b) represents a steady arrival of medium input, long output, where a batch of nearly 9 requests is sent every 3 seconds.}
    \label{fig:requests}
\end{figure}

\begin{figure*}[t]
    \centering
    \includegraphics[width=\linewidth]{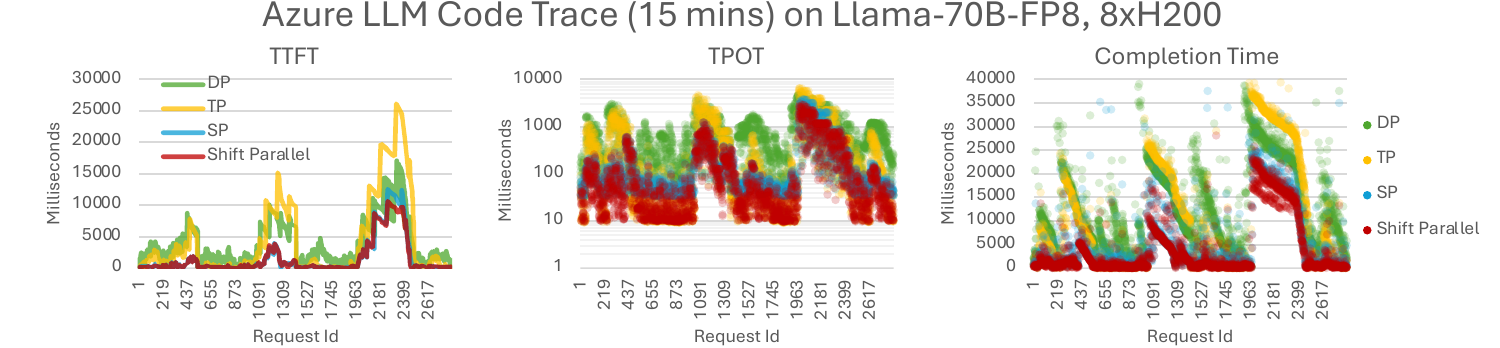}\\
    \caption{The Azure trace has three prominent bursts (see requests 437, 1091, 2181) and those coincide with the bursts experiencing a higher TTFT and TPOT, and hence the completion time increases in during the bursts. The requests at the beginning of a burst experience a higher completion time because their decode overlaps with the prefill of consecutive requests.}
    \label{fig:real_world_traces_azure}
\end{figure*}

\begin{figure*}[t]
    \centering
    \includegraphics[width=\linewidth]{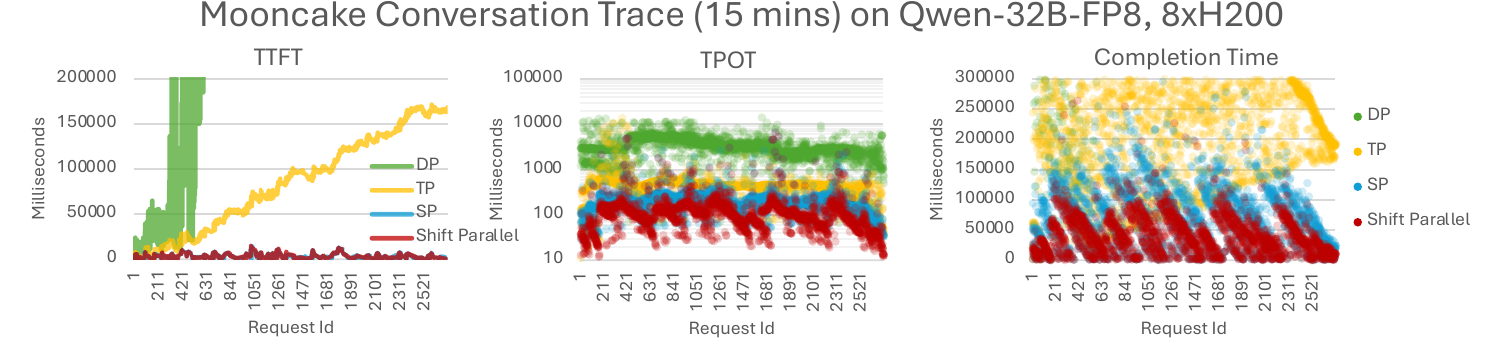}\\
    \caption{The Mooncake trace involves a heavier workload. DP and TP cannot keep up with the traffic, and cause wait times to grow indefinitely as seen in TTFT. SP and Shift Parallel can sustain the conversation traffic with a finite completion time.}
\label{fig:real_world_traces_mooncake}
\end{figure*}

\subsubsection{Real-Life Workload Traces}\label{sec:real_world_datasets}


In this section, we study two real-life traces. We performed these case studies on code completion and conversation on a single node based on the following open-source traces: 1) Azure LLM Code Trace \cite{patel2024splitwise}, and 2) Mooncake Conversation Trace \cite{qin2025mooncake}. For demonstration we run both traces for 15 minutes. The resulting request distribution and arrival rates are shown in Figure~\ref{fig:requests}.

\paragraph{Azure LLM Code Trace on Llama-70B}
This trace is produced on Azure platform from real-world agentic code generation. We replay these traces on our platform (Section~\ref{sec:hardware}) with the Llama-70B-FP8 model.

As seen in Figure~\ref{fig:real_world_traces_azure}, DP (shown with green) is throughput oriented, and handles the bursts better compared to TP (shown with yellow), yielding a lower TTFT and completion time in high traffic. TP is latency oriented and yields a lower TPOT, and hence completion time than DP in low traffic.

\begin{figure}[t]
    \centering
    \includegraphics[width=\linewidth]{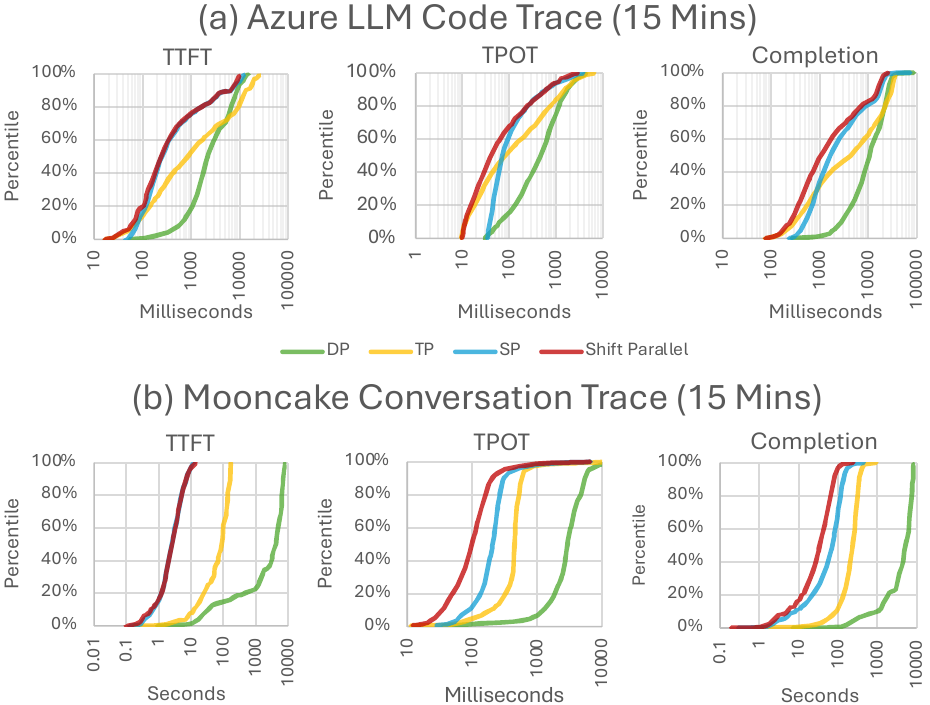}
    \caption{Request latency distributions in the (a) Azure (b) Mooncake traces. Shift Parallelism is more likely to deliver the lowest completion time in either case.}
    \label{fig:real_world_pdf}
\end{figure}

Shift Parallelism (Figure~\ref{fig:real_world_traces_azure}, red) is efficient across traffic and obtains the lowest TTFT, TPOT, and hence completion time. The statistics in Figure~\ref{fig:real_world_pdf} (a) show that shift parallelism helps to achieve tighter service-level objectives (e.g., p50, p99), compared to other parallelisms.

\paragraph{Mooncake Conversation Trace on Qwen-32B}
This trace is from real-world conversation on Moonshot AI’s platform. In this case, we chose to use a smaller model (Qwen-32B-FP8) because the Llama model did not sustain the traffic and context size in our platform, i.e., the arrival rate is so high for a single node that KV cache becomes full, causing wait times as seen in DP and TP in Figure~\ref{fig:real_world_traces_mooncake}.

In our initial try, Qwen-32B-FP8 did not sustain the conversation traffic. Therefore we turned on FP8 KV cache data type (originally it is FP16) for increasing the KV cache capacity. The wait times were reduced, yet DP and TP still could not sustain the incoming traffic yet Shift Parallelism could run the trace without KV cache overflow. Therefore only SP and Shift Parallel solutions can support this trace on a single node without creating significant queuing delays. The statistics are shown in Figure~\ref{fig:real_world_pdf} (b).

\begin{figure*}[t]
    \centering
    \includegraphics[width=\linewidth]{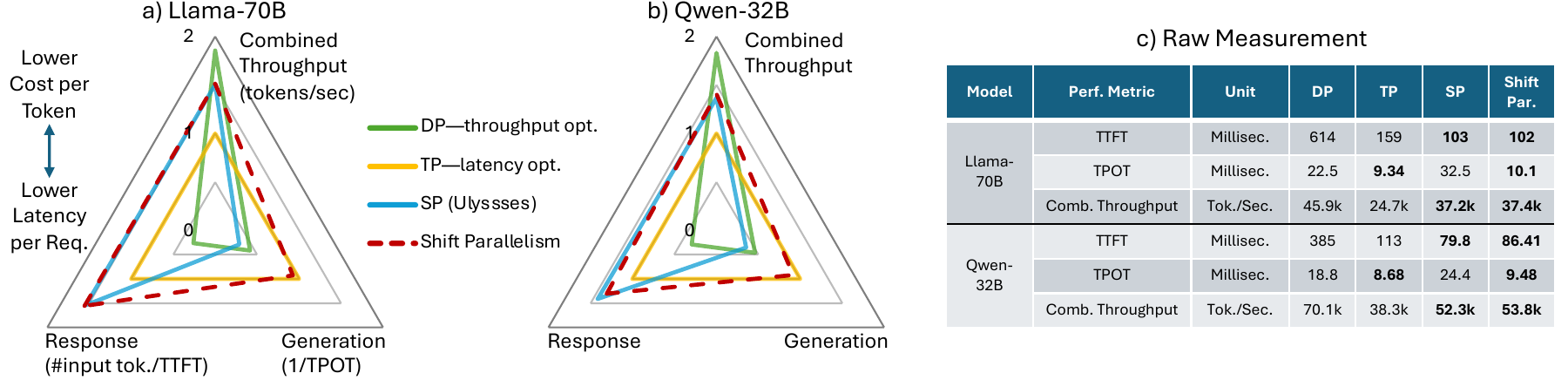}
    \caption{Comparison of response and generation latency, and throughput, all in tokens/sec. with (a) Llama-70B and (b) Qwen-32B based on the (c) measurements. Shift Parallelism simultaneously obtains a higher throughput and a lower latency than TP, alleviating the latency vs. throughput tradeoff of existing parallelisms.}
    \label{fig:trifecta_llama_qwen}
\end{figure*}

\begin{figure*}[b]
    \includegraphics[width=\linewidth]{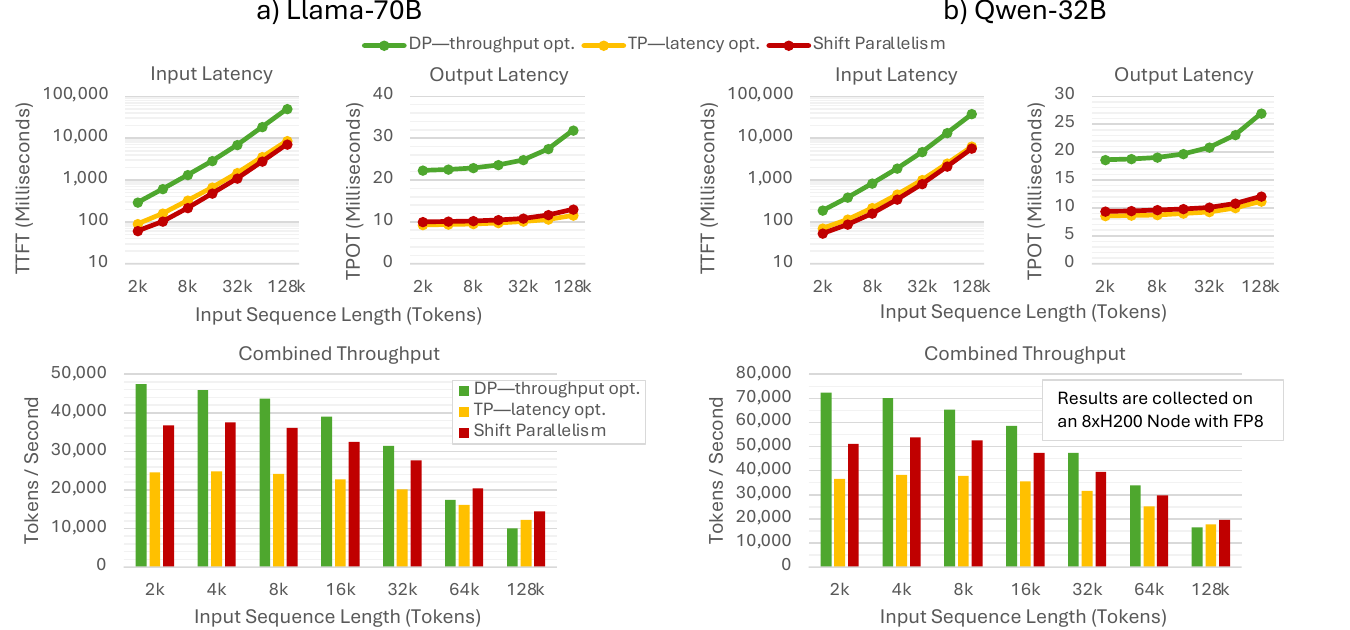}
    \caption{Performance variation across input sequence length: Minimum latency (TTFT, TPOT), and maximum throughput of (a) Llama-70B and (b) Qwen-32B. The throughput drops with large context sizes due to the excess attention time.}
    \label{fig:input_size}
\end{figure*}

\subsection{Performance Benchmarks}\label{sec:performance analysis}

In this section we send synthetic requests for presenting performance characteristics via parameterized experiments.

\subsubsection{Latency vs. throughput tradeoff}

We evaluate the latency vs. throughput tradeoff across parallelism using a uniform request size of 4k input tokens and 250 output tokens. For finding the peak throughput, we send a batch of requests (thousands) and provide sufficient concurrency to saturate the GPU throughput. For finding the lowest latency, we process requests sequentially, i.e., a single request at a time.

Figure~\ref{fig:trifecta_llama_qwen} compares DP, TP, SP, and Shift Parallelism across Llama-70B and Qwen-32B. Shift Parallelism achieves the lowest TTFT that is 1.56$\times$ and 6$\times$ lower than TP and DP for Llama, and 4.45$\times$ and 1.31$\times$ for Qwen. Shift Parallelism achieves that lowest TPOT that is 9.34 ms for Llama and 8.68 ms for Qwen. Shift Parallelism experiences significantly less throughput degradation compared to TP. Specifically, TP loses 46\% and 45\% throughput with Llama and Qwen, yet Shift Parallelism only loses 18\% and 23\%, respectively. A bit part of it comes from vLLM's engine overhead that we discuss in Section~\ref{sec:breakdown}.

\subsubsection{Variations across context sizes.}

For investigating TTFT, TPOT, and throughput with various input context sizes, we repeat the experiment in Figure~\ref{fig:trifecta_llama_qwen} for input sequences with 2k--128k tokens and 250 output tokens. Figure~\ref{fig:throughput_latency} presents the input and output latency in low traffic, and throughput in high traffic.

Shift Parallelism provides a 6.97$\times$ and 1.56$\times$ faster response than DP and TP, respectively, because it uses SP for prefill, which is more efficient than both DP and TP. As a result, Shift Parallelism is more responsive and also provides a faster completion time especially for long input and short output contexts where TTFT dominates the completion time (such as in summarization).

Shift Parallelism provides up to 2.45$\times$ faster generation than DP and a similar output latency to TP. Ideally, the output latency should not depend on the input context size in theory, but in practice TPOT increases with the input size (see Figure~\ref{fig:input_size}). The main reason is that each output token needs to read more number of tokens from the KV cache as input context grows, and eventually the system becomes memory bandwidth bound. TP and Shift Parallelism parallelize the attention layer (Section~\ref{sec:transformer}), and hence the KV cache, providing memory bandwidth, mitigating the output latency for long inputs.

Shift Parallelism obtains up to 1.51$\times$ higher peak throughput than TP, meaning that processing the high-traffic bursts and also batch workloads is approx. 50\% faster with Shift Parallelism. Nevertheless, the throughput drops significantly with larger contexts because attention time dominates the end-to-end generation. See Section~\ref{sec:breakdown} for analysis.

\subsubsection{Latency vs. Arrival Rate}

%



To investigate the performance between extremely high and low traffic rates, we test Shift Parallelism across a wide range of intermediate traffic by varying the request arrival rates. We measure TTFT and TPOT of an individual request, which both increases with higher traffic, and calculate the completion time as $\textrm{TTFT} + \textrm{\#output tok.} \times \textrm{TPOT}$.
The question is, where does the tradeoff happen, and how well Shift Parallelism transitions from latency to throughput optimization?

\begin{figure}[h]
    \centering
    \includegraphics[width=\linewidth]{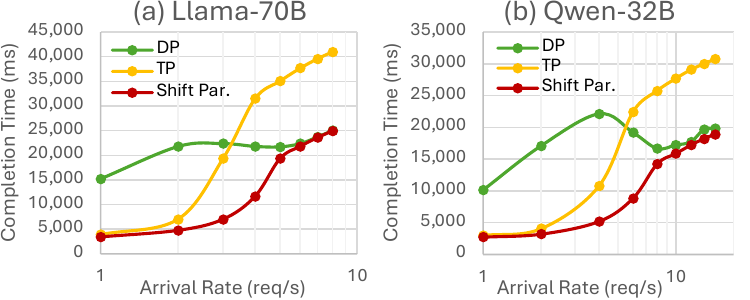}
    \caption{Request completion time vs. arrival rate. TP and DP make the performance tradeoff across arrival rates. Shift Parallelism strictly obtains the lowest completion time across arrival rates. Request size: 8k input, 250 output.}
    \label{fig:arrival_rate}
\end{figure}

Results in Figure~\ref{fig:arrival_rate} shows the performance variation across arrival rates. TP and DP curves cross over at a critical arrival rate (a few req/sec). Yet Shift Parallelism guarantees the lowest latency regardless of the arrival rate---strictly better than both DP and TP solutions. In low-to-medium rates (req/s), Shift Parallelism switches back-and-forth across SP and TP for minimizing the input (TTFT) and output (TPOT) latencies, respectively. In high traffic, Shift Parallelism uses SP to save combined throughput (tokens/sec).

\subsection{Cost Breakdown}\label{sec:breakdown}
\begin{figure}[t]
    \centering
    \includegraphics[width=\linewidth]{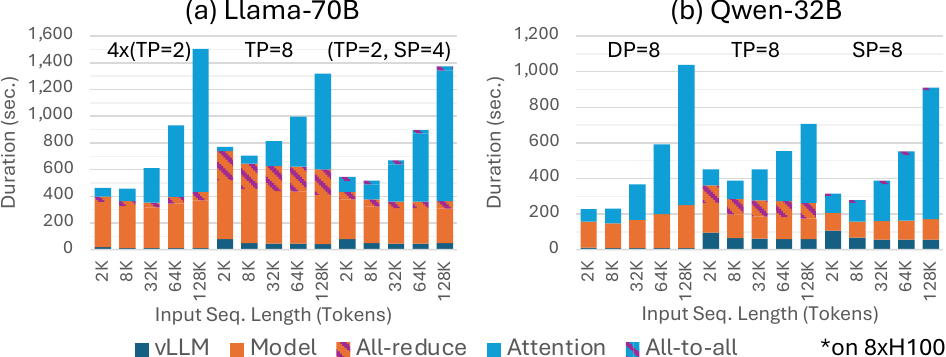}
    \caption{End-to-end cost breakdown* of time spent in a batch workload with (a) Llama-70B and (b) Qwen-32B. Shorter seq. $\rightarrow$ vLLM overhead, longer seq. $\rightarrow$ attn. time.}
    \label{fig:breakdown}
\end{figure}

We analyze the cost of individual system components by taking away one component at a time. Figure~\ref{fig:breakdown} shows the resulting breakdown of time to process a batch of requests with Llama-70B and Qwen-32B models on a single node. We clearly see that SP (and hence Shift Parallelism) has a lower communication cost than TP. On the other hand, we observe two unaddressed performance bottlenecks that are not related to Shift Parallelism:
\begin{itemize}[left=0pt]
    \item Attention time grows significantly with the sequence size, and therefore reduces the combined throughput. Recent papers address this issue using sparse attention~\cite{Child2019GeneratingLS} and it is out of scope of this paper.
    \item The parallelization cost of vLLM is significant in small models (e.g., compare Llama-70B, Qwen-32B). We find vLLM cost by removing the forward pass. This indicates that a large portion of the remaining throughput gap between DP and SP might actually be the vLLM overhead unrelated to SP.
\end{itemize}

\subsection{Shift Parallelism in Production}\label{sec:comparison}

We fully integrated Shift Parallelism in our existing production environment (see Section~\ref{sec:integration}). Running efficiently in production is not only about parallelism, but it also requires a plethora of other state-of-the-art techniques. To that extent, we integrated  Shift Parallelism with SwiftKV~\cite{qiao2025swiftkvfastprefilloptimizedinference} and speculative decoding~\cite{arctic-speculator,oliaro2025suffixdecodingextremespeculativedecoding} in our production environment~\cite{rajbhandari2025arcticinferenceshiftparallelism}.  

\begin{figure}[b]
    \centering
\includegraphics[width=\linewidth]{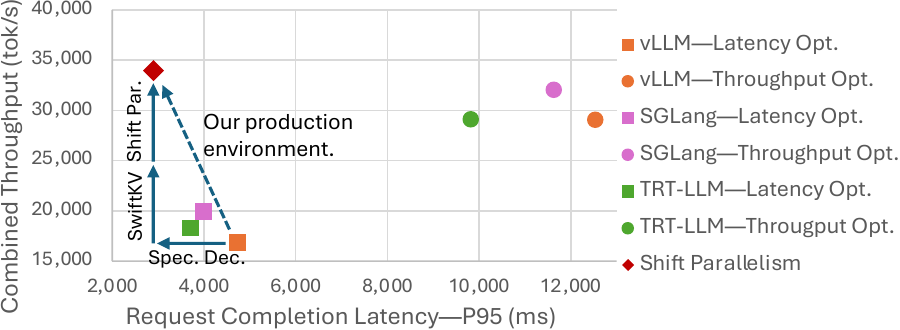}
    \caption{Compounding effect of optimizations and comparison of our production environment with other frameworks out-of-the-box using a real dataset on Llama-70B. Latency- and throughput-optimized  configurations use TP and DP, respectively. }
    \label{fig:comparison}
\end{figure}

Figure~\ref{fig:comparison} shows that our combined production simultaneously achieves highest throughput (lowest cost) and lowest completion time---all in one deployment---outperforming the best open source systems optimized for each metric individually
\footnote{We enabled the best available speculative decoding for each framework. These experiments were run on data sets generated using real-world production traces to compute throughput, and a mixture of ShareGPT, HumanEval and SWEBench to measure latency. As a result, these results are representative of performance achievable in real-world deployments.}. The compounding effect of SwiftKV and speculative decoding are denoted. 



\subsection{Limitations and Future Work}

To investigate further on Shift Parallelism's performance behavior, we stretch our evaluation to two recently released sparse (i.e., MoE) models that are listed in the bottom two rows of Table~\ref{tab:models}.

To enable these models, we use the SP generalizations in Section~\ref{sec:design}: i) Llama-17B-16E barely fits into a single GPU and when $SP=8$ is used in the base config., there is no memory left in the KV cache to support large context sizes. To enable long contexts, we use Algorithm~\ref{alg:combined_SP_TP} for the base config. $(SP=4, TP=2)$. ii), Qwen-30B-A3B suffers from scaling because it only has 4 KV heads, and we use KV cache replication (Section~\ref{sec:sp_inference}) to scale the model across $SP=8$ GPUs.

\begin{figure}[t]
    \centering
    \includegraphics[width=\linewidth]{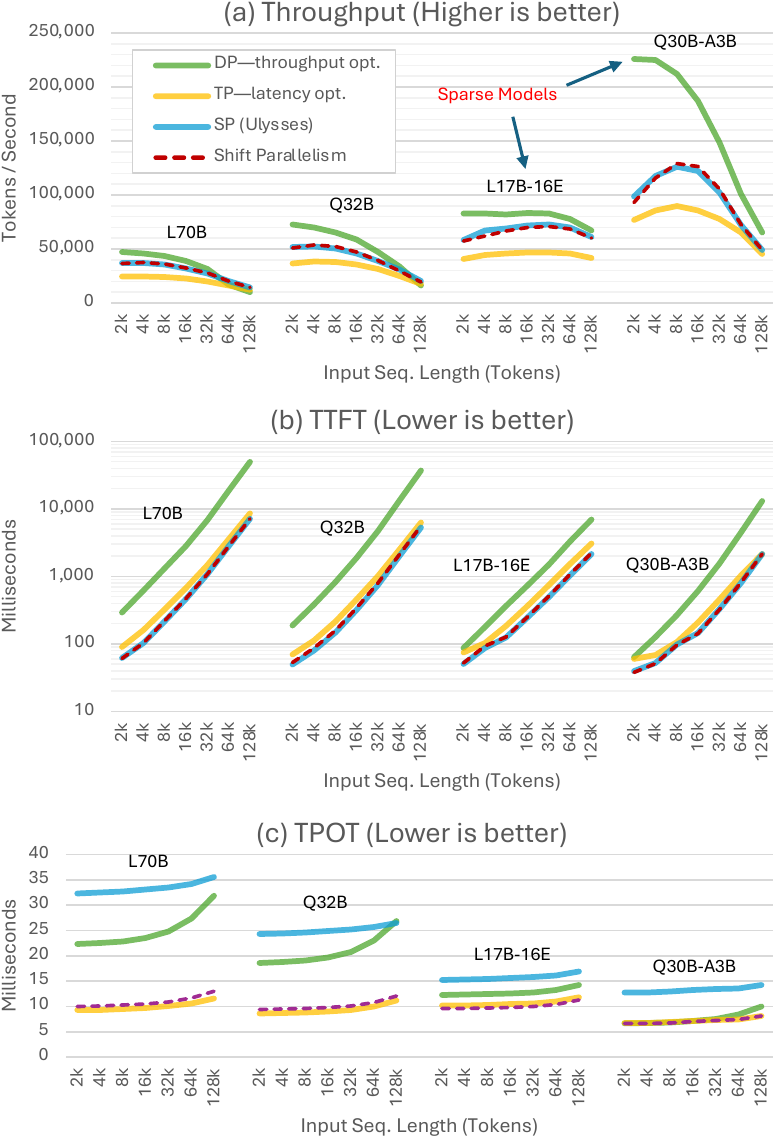}
    \caption{(a) Peak Throughput and minimum latency---(b) TTFT and (c) TPOT---comparison across parallelisms, models and input sequence lengths. Shift Parallelism switches across SP and TP for obtaining high throughput and low latency. 
    }
    \label{fig:throughput_latency}
\end{figure}

Figure~\ref{fig:throughput_latency} compares the throughput and latency performance in across the board. The models sorted from larger to smaller. The sparse models attain a higher throughput and lower latency than dense models, simply because they have fewer active parameters (see Table~\ref{tab:models}). Compared to TP, Shift parallelism shows excellent performance on both sparse models: Up to 50\% higher throughput without increasing the latency. 

The smallest model, Qwen-30B-A3B, attains a much higher throughput (225k tokens/sec.) than the other models with DP, and the throughput suddenly drops either with TP and SP, especially with smaller end of input sizes. The discrepancy is due to the vLLM's parallelization overhead with small models that is discussed in Section~\ref{sec:breakdown}.

The sparse models beg further investigation in the context of expert parallelism (EP). Specifically, there is no prior work that combines SP with EP to further optimize sparse models, which we will leave as a future work.

\section{Related Work}\label{sec:related}

The heterogeneous resource demands of LLM inference is one of its main challenges, and prior works have proposed different techniques to manage it. We discuss a few works related to Shift Parallelism in this section.

\emph{Chunked Prefill}~\cite{agrawal2024tamingthroughputlatencytradeoffllm,holmes2024deepspeedfastgenhighthroughputtextgeneration} tackles the heterogeneous resource demand of LLM inference requests in their input-processing (prefill) stage vs their output-generation (decode) phase. During prefill, the input tokens are processed in parallel and is thus compute-intensive, while during decode, a single token is generated at a time but requires the full KV-cache of prior tokens, and is thus memory-intensive. Chunked prefill is proposed by DeepSpeed-FastGen~\cite{holmes2024deepspeedfastgenhighthroughputtextgeneration} and Sarathi-Serve~\cite{agrawal2024tamingthroughputlatencytradeoffllm} and mixes requests in both prefill and decode phases in the same batch, thereby improving resource utilization and combined throughput. Today, it is default in many popular inference engines~\cite{kwon2023efficientmemorymanagementlarge,zheng2024sglangefficientexecutionstructured,nvidia_tensorrt_llm_2023}. In comparison, Shift Parallelism is targeted at time-varying sizes of each batch. It is orthogonal and compatible with chunked prefill, and our experiments in Sec.~\ref{sec:evaluation} are run using their combination.

\emph{Disaggragated Inference}~\cite{zhong2024distservedisaggregatingprefilldecoding,qin2024mooncakekvcachecentricdisaggregatedarchitecture} uses separate GPU workers for prefill and decode so that prefill throughput and decode latency can be separately scaled and optimized. Compared with chunked-prefill systems and Shift Parallelism, disaggregated inference can eliminate interference between requests in the prefill and decode stages, but at the cost of dedicating additional resources to each stage. Additionally, the KV cache must be communicated from a prefill worker to a decode worker for each request, causing extra communication overhead. In contrast, Shift Parallelism with chunked-prefill overlaps prefill and decode, with decode tokens accessing the KV cache from local memory, resulting in more efficient resource utilization and less cost per token.

\section{Conclusion}\label{sec:conclusion}

LLM's have diverse real-world applications that yield different traffic patterns that have different performance requirements. Fundamentally, existing parallelisms (TP and DP) for inference optimize either for latency (for interactive applications) or throughput (for batch workloads), but do not optimize for multiple applications in the same deployment.

For supporting dynamic workloads, we need to switch back-and-forth between parallelisms, yet cannot switch across TP and DP due to their KV cache mismatch. As a remedy, we bring in SP (Ulysses) that is originally applied to training for high throughput, and generalized it for inference, and providing the KV cache invariance across SP and TP  by addressing corner cases. We call our solution Shift Parallelism, where a deployment has two configurations. that switch back and forth across SP (base config.) and TP (shift config.).

Our extensive benchmarking shows that Shift Parallelism addresses the latency vs. throughput tradeoff in a low cost way, providing up to 50\% more throughput compared without losing latency across traffic rates. We fully integrated Shift Parallelism along with other SoTA that we use in production, and open source our inference system. 

\bibliographystyle{ACM-Reference-Format}
\bibliography{shapeshifter}

\appendix
\section{Artifact Appendix}

\subsection{Abstract}

Our evaluation in Section~\ref{sec:evaluation} involves extensive results. For assessing the real-world benefit, we will focus on running open-source production traces (Section~\ref{sec:real_world_datasets}). The artifact compares vLLM's built-in parallelisms (DP and TP) and the proposed parallelisms (SP and Shift Parallelism) using \texttt{ArcticInference} plug-in system that we also use in our production. For reproducibility, we included the instructions in the repository, and uploaded the cleaned traces into a permanent archive.

\subsection{Artifact check-list (meta-information)}

{\small
\begin{itemize}
  \item {\bf Algorithm: } Multi-GPU parallelism strategies (DP, TP, SP, Shift Parallel) for LLM inference.
  \item {\bf Program: } vLLM v0.10.1 and the corresponding ArcticInference plug-in version (both are open source).
  \item {\bf Model: } We use two open-source models downloadable from Hugging Face.
  \item {\bf Data set: } Two open-source production traces. Clean traces for reproducibility are archived in DOI: \\
  https://doi.org/10.5281/zenodo.18240909.
  \item {\bf Run-time environment: } Linux, Python 3.10
  \item {\bf Hardware: } A node with 8$\times$H200 Nvidia GPUs w/ NVLinks.
  \item {\bf Metrics: } TTFT, TPOT, completion time
  \item {\bf Output: } vLLM's benchmarking script prints the result.
  \item {\bf Experiments: } Running vLLM benchmarking scripts (server and client). We provide the scripts.
  \item {\bf How much disk space required (approximately)?: } 100 GB to store both models. 
  \item {\bf How much time is needed to prepare workflow (approximately)?: } 15 mins
  \item {\bf How much time is needed to complete experiments (approximately)?: } 5 node-hours. Experiments can be easily done in parallel across two nodes.
  \item {\bf Publicly available?:}\\ \url{https://github.com/snowflakedb/ArcticInference}
  \item {\bf Code licenses?: } Apache-2
\end{itemize}
}

\subsection{Description}

\subsubsection{How to access}

vLLM is open source (\url{https://github.com/vllm-project/vllm}).

\subsubsection{Hardware dependencies}

We used the hardware described in Section \ref{sec:hardware}), yet a generic Nvidia DGX-H200 node is sufficient for 1:1 reproduction. Optimal configurations, and hence the results may look different another type of multi-GPU node, yet the conclusion should be the same.

\subsubsection{Software dependencies}

We depend on \texttt{vLLM v0.10.1} and the corresponding \texttt{ArcticInference} plug-in (commit \texttt{5e08f0f}).

\subsubsection{Data sets}

We use the following datasets (links given):
\begin{itemize}
    \item \href{https://github.com/Azure/AzurePublicDataset/blob/master/data/AzureLLMInferenceTrace_code.csv}{Azure LLM code}
    \item \href{https://github.com/kvcache-ai/Mooncake/blob/main/FAST25-release/traces/conversation_trace.jsonl}{Mooncake conversation}
\end{itemize}

We provide the cleaned traces (15 Mins each) at DOI: https://doi.org/10.5281/zenodo.18240909 that reproduces Figures~\ref{fig:real_world_traces_azure}--\ref{fig:real_world_traces_mooncake}. We also include a vLLM benchmarking patch for running these traces in our repository.

\subsubsection{Models}

The models are open-source in Hugging Face (links given):
\begin{enumerate}
    \item \href{https://huggingface.co/RedHatAI/Llama-3.3-70B-Instruct-FP8-dynamic}{\texttt{RedHatAI/Llama-3.3-70B-Instruct-FP8-dynamic}}
    \item \href{https://huggingface.co/Qwen/Qwen3-32B-FP8}{\texttt{Qwen/Qwen3-32B-FP8}}
\end{enumerate}
These can be downloaded as \texttt{huggingface-cli download <model> --local-dir <model>}

\subsection{Installation}

We built our work on vLLM, therefore it is essential to make vLLM work first. The installation details can be found in \url{https://github.com/vllm-project/vllm}, yet it should be as easy as: \texttt{pip install vllm==v0.10.1}

\subsection{Experiment workflow}

The step-by-step instructions can be found in \\\texttt{ArcticInference/benchmark/reproducibility}.First the output file of an experiment should be obtained. Then the plotting script should be used to plot TTFT, TPOT, and completion time. The scripts for running the experiments and plotting them as in Figure~\ref{fig:real_world_traces_azure}--\ref{fig:real_world_traces_mooncake} are provided.

\subsection{Evaluation and expected results}

Shift parallelism is expected to have a lower TTFT, TPOT, and completion compared to TP and DP. The raw files can be plotted using \texttt{plot.py}. Additionally, the columns (TTFT, TPOT, Completion) can be sorted to obtain Figure~\ref{fig:real_world_pdf}.






\end{document}